\documentclass[5p,12pt,twocolumn]{elsarticle}

\usepackage{amsmath}              
\usepackage{epsfig}                   
\usepackage{hyperref}               

\makeatletter
\newcommand{\Rmnum}[1]{\expandafter\@slowromancap\romannumeral #1@}   
\makeatother

\begin{document}

\begin{frontmatter}

\title{Ca-intercalated graphite as a hydrogen storage material: stability against decomposition into CaH$_{2}$ and graphite}

\author[ucl,lcn]{C.R.~Wood\corref{cor1}}
\ead{c.wood@ucl.ac.uk}
\author[ucl,lcn]{N.T.~Skipper}
\author[ucl,lcn]{M.J.~Gillan}

\cortext[cor1]{Corresponding author}

\address[ucl]{Department of Physics and Astronomy, University College London, Gower Street, London, UK WCIE 6BT}
\address[lcn]{London Centre for Nanotechnology, University College London, 17-19 Gordon Street, London, UK WC1H 0AH}

\begin{abstract}
We have used calculations based on density functional theory to investigate the energetics of hydrogen absorption in calcium-intercalated graphites. We focus particularly on the absorption energy and the stability of the hydrogenated material with respect to decomposition into graphite and calcium hydride, which is essential if this material is to be useful for practical H$_{2}$ storage. The calculations are performed with two commonly used approximations for the exchange-correlation energies. Our calculations confirm earlier predictions that the absorption energy is approximately $-0.2$ to $-0.4$~eV, which is favourable for practical use of Ca-intercalated graphite as a hydrogen storage medium. However, we find that the hydrogenated material is strongly unstable against decomposition. Our results therefore explain recent experiments which show that H$_{2}$ does not remain stable in CaC$_{6}$ but instead forms a hydride plus graphite.
\end{abstract}


\end{frontmatter}

\section{Introduction}  %
\label{sec:introduction}%

Finding an alternative, or secondary, energy source to replace petroleum for use in vehicles is of paramount importance. Hydrogen is a possible candidate and, with the advancement of proton exchange membrane (PEM) fuel cell technologies \cite{Mehta_2003}, could provide a viable energy solution. The practical ambient storage of hydrogen has long been a challenge \cite {Zhou_2005, Harris_2004}. If hydrogen fuel cells are to be used in the transport industry then a lightweight energy dense storage medium needs to be found \cite{Ross_2006}. The U.S. Department of Energy state that a storage medium needs to store 6.5$\%$ \cite{DoE_2009} hydrogen by weight at ambient temperature and pressure. In order that the material can take in and release hydrogen at ambient temperature and reasonable pressures, the absorption energy also needs to be in the range $0.2-0.8$~eV ($1.000$~eV $=96.485$~kJ/mol). 

Graphite intercalation compounds (GICs) have shown promise as a potential ambient high-density storage medium because they have a low density. They have the great advantage that their chemistry and the spacing between the carbon planes can be tuned \cite{Dresselhaus_1980, Lovell_2006}. They are also inexpensive to produce. Many GICs have been considered for hydrogen storage and possess very different properties when hydrogen is introduced into them \cite{Lovell_2008, Carlile_1998, Beaufils_1981}. Recently calculations based on density functional theory (DFT) have been used to survey the binding energies of hydrogen in a wide range of metal-intercalated graphite materials \cite{Cobian_2008}. The results showed that alkaline-earth intercalates appear to be very promising, because their calculated binding energies are in the required range. An essential requirement for the practical use of any of these GIC materials is that they should be stable against decomposition into the metal hydride and graphite. Very recent experiments \cite{Srinivas_2009} on the hydrogenation of calcium graphite (CaC$_{6}$) suggest that this may be a major issue and that the attempted hydrogenation of the material leads to the formation of orthorhombic calcium hydride (CaH$_{2}$) and the complete deintercalation of the starting CaC$_{6}$ leaving graphite. This has also been shown to occur for lithium graphite (LiC$_{6}$) \cite{Enoki_1990}. Surprisingly, this crucial question of stability against decomposition does not appear to have been addressed before by modelling. The recent DFT work by \citet{Cobian_2008} on hydrogen storage in GIC materials did explore the question of stability against metal clustering, but did not consider the possibility of hydride formation. 

The purpose of the present paper is to use electronic-structure methods to study the stability of hydrogenated Ca-intercalated graphite against decomposition into graphite and CaH$_{2}$. The experiments of \citet{Srinivas_2009} indicate that this decomposition occurs spontaneously, so that it must be exothermic. This means that the Ca-GIC could not be used without modification as a hydrogen storage material. However, it is still possible that the hydrogenated material might become stable against decomposition if other atomic or molecular species were co-intercalated together with Ca. In considering whether and how this can be done, it is important to know the heat of decomposition of hydrogenated Ca-GIC. This important quantity is not available from the experiments, and our aim here is to use DFT to calculate it. 

In the next section, we summarise the essential background information about Ca-GICs, and we also outline the DFT techniques used in this work. In section \ref{sec:results}, we present our calculations, first on Ca-GIC (\ref{subsec:CaCn}) itself at various stoichiometries, then on the hydrogenated material (\ref{subsec:H2 in CaCn}), and finally on the CaH$_{2}$ crystal (\ref{subsec:CaH2}). These calculations yield values of the heat of decomposition of hydrogenated Ca-GIC, computed with different DFT approximations. As expected, the decomposition is exothermic, and we discuss in section \ref{sec:conclusion} the implications of our results.

\section{Background information}  %
\label{sec:background}                   %

\subsection{The materials}
\label{subsec:materials}

The structure of graphite is well known to consist of planes of carbon atoms, with each plane having a hexagonal honeycomb structure. Different registrations of the carbon planes relative to each other are possible, the main two being $A-B-A$ (hexagonal, 2H) and $A-B-C$ (rhombohedral, 3R), with the 2H registration being most thermodynamically stable at standard temperature and pressure \cite{Shi_1997}. The nearest neighbour C-C distance is 1.42~\AA. The layers are held together by weak van der Waals dispersion, a physical effect that is not included in standard forms of DFT. Nevertheless, it is known that the local density approximation (LDA) of DFT fortuitously predicts an equilibrium interplanar spacing (3.38~\AA) that is very close to the experimental value (3.36~\AA). 

\begin{figure}[ht]
 \begin{center}
   \includegraphics[width=7.5cm]{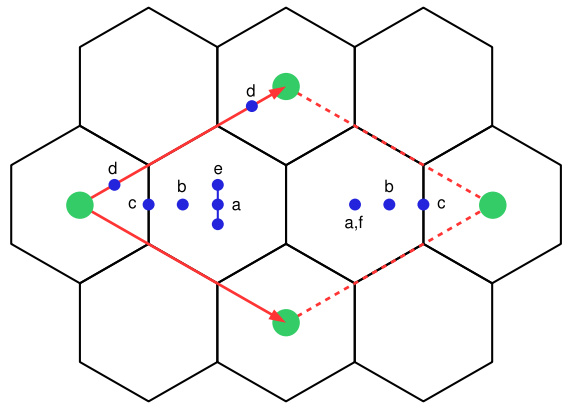}
  \end{center}
  \caption{The 6 starting positions of hydrogen in CaC$_{6}$. Note positions (\emph{e}) and (\emph{f}) are both H$_{2}$ molecules with (\emph{f}) perpendicular to the page.}
  \label{fig:H+H+CaC6_start_posn}
\end{figure}

For Ca-intercalated graphite the registrations of the graphite layers are shifted so that carbon atoms superimpose each other producing an $A-A-A$ stacking. It has been shown by x-ray diffraction experiments \cite{Emery_2005} that the Ca intercalants are positioned directly above and below hexagon centres (see figure \ref{fig:H+H+CaC6_start_posn}). The only experimentally observed stoichiometry is CaC$_{6}$ in $\alpha-\beta-\gamma$ stacking (where we use the notation of \citet{Dresselhaus_1980} with Greek letters labelling successive intercalant layers) such that each successive Ca layer is shifted relative to the preceding one resulting in a rhombohedral structure. This only exists as a \textit{stage \Rmnum{1}} GIC, where the staging numbers how many graphite layers exist between each intercalant layer. It is known that other GICs produce different stage GICs, in the case of KC$_{24}$ a stage \Rmnum{2} GIC. The interplanar spacing of Ca-GIC is increased with respect to graphite to 4.524~\AA$\;$ (a 35$\%$ increase over graphite) and the C-C in-plane nearest neighbour distance to 1.444~\AA$\;$ (a small 1.7$\%$ increase over graphite). 

A definable crystal structure for the hydrogenated Ca-GIC is not observed experimentally \cite{Srinivas_2009a}. Once hydrogenated the CaC$_{6}$ diffraction peaks weaken in intensity and the reflections from the CaH$_{2}$ structure become apparent. A weak unidentified peak appears but is too weak to relate to a major structure change. 

The CaH$_{2}$ crystal has an orthorhombic unit cell with $Pnma$ symmetry (International Tables group 62). The unit cell consists of four CaH$_{2}$ units at each of the Wyckoff 4c positions. This means that the crystal requires only the x- and z-coordinates of one Ca atom and of two inequivalent H atoms to fully describe it; the y-coordinate is fixed by symmetry. The Wyckoff 4c positions are given in table \ref{tab:WyckoffPosns}. Experimental structural parameters of the crystal have been reported by \citet{Wu_2007}, and calculations on the structure have been performed by \citet{Li_2007} and \citet{Wolverton_2007}. 

\subsection{Electronic-structure techniques}
\label{subsec:electronic techniques}

This work employs standard electronic-structure methods based on DFT, applied to periodically repeated simulation cells. At the temperature of interest, electronic excitations are negligible, and all the predictions are derived from the electronic ground state energy of the system and the forces on the ions for given ionic positions. We use the projector augmented wave (PAW) method to determine the electronic ground state, and all the calculations are performed with the VASP code. We have always made efforts to ensure that the calculations are well converged with respect to the PAW plane-wave cutoff, so as to ensure basis-set completeness, and $k$-point sampling to reduce or eliminate errors due to the finite size of the simulation cell. Details of the plane-wave cutoff and $k$-point sampling will be given in the results section, where appropriate.

The accuracy of DFT predictions depends mainly on the approximation used for the exchange-correlation functional. This is an important issue in the present work, since the adsorption energy of hydrogen in Ca-GIC is fairly small - only a few hundred meV, and DFT errors are likely to be significant on this scale. To gauge the likely errors, we have performed the calculations with two rather different functionals: the LDA and the PW91 form of GGA. There is a general tendency for GGA functionals to be more accurate for binding energies, but their failure to produce any binding between the layers of pure graphite is a significant deficiency in the present context. On the other hand, LDA usually over binds, but its good prediction of the equilibrium interplanar spacing in graphite may be helpful here. The differences between LDA and PW91 predictions can therefore be used as a rough indication of the likely DFT errors.

When searching for relaxed minimum energy structures, we generally use the automatic optimisation algorithms of VASP to relax the ionic positions. The VASP algorithms can also be used to relax the size and shape of the simulation cell, but we found that it is often helpful to relax the cell parameters by hand, and we will note how this has been done at the appropriate places in the results section. 

\section{Results}  
\label{sec:results}

\subsection{Pure graphite}
\label{subsec:pure graphite}

As a check, we have repeated the calculations performed by earlier workers on pure graphite in the 2H, $A-B-A$ structure that is observed experimentally. We found that the LDA and PW91 predictions of the in-plane C-C bond length are \textit{a}=1.41~\AA $\;$  and 1.43~\AA $\;$  respectively, which are very close to the experimental value of 1.42~\AA $\;$  \cite{Shi_1997}, while LDA gives the value of \textit{c}=3.38~\AA $\;$ , in close agreement with the experimental value of 3.36~\AA $\;$  \cite{Shi_1997}. Our calculations on the $A-A-A$ structure give the same values for \textit{a} as in the 2H structure, and LDA gives \textit{c}=3.60~\AA $\;$ in this registration. 

\subsection{Calcium intercalated graphite (CaC$_{n}$)}
\label{subsec:CaCn}

As noted above, we only need to study stage \Rmnum{1} Ca-GICs here. We shall present results for the different stackings $\alpha$, $\alpha \beta$ and $\alpha \beta \gamma$ for CaC$_{6}$ but since the structural parameters and total energies turn out to be essentially identical, we examine only $\alpha$ stacking for other stoichiometries. The possible stoichiometries for CaC$_{n}$ are given by the formulae
\begin{eqnarray}
  n & = & \left\{
\begin{array}{l} 
  2\left(k\left(k+2\right)+1\right) \\ 
  2\left(k\left(k-1\right)+1\right)
\end{array}
\right.,  
\end{eqnarray}
with $k=1,2,3\ldots$, a positive integer. Here, we consider the cases $k=1,2,3$, which correspond to $n=6,8,14$. 

The unit cell was relaxed for all the Ca-GICs and the automatic algorithm in VASP used to relax the atomic positions within this equilibrium cell. A plane-wave basis set energy cutoff of 800~eV and a $k$-point grid with spacings always smaller than 0.04~\AA$^{-1}$ were used for all calculations.

The two parameters that fully specify the structure are the C-C in-plane bond length \textit{a} and the inter-planar spacing \textit{c}. Our calculated values of \textit{a} and \textit{c} are reported in table \ref{tab:CaCncellparameters}, which also includes the available experimental data. We note that for CaC$_{6}$, the differences between the different stackings are extremely small. The value of \textit{a} is greater than that of graphite by only 1.5pm (1.0\%) with PW91 and 2.5pm (1.8\%) with LDA. The predicted values of \textit{c} obtained with LDA and PW91 are fairly close to each other, and the latter is in excellent agreement with experiment. The value of \textit{c} is greater than that of graphite by $\sim$34\%.

\begin{table*}[htp]
  \caption{Calculated unit cell parameters for the CaC$_{n}$ graphite intercalate and the comparison of CaC$_{6}$ $\alpha \beta \gamma$-stacking with experiment \cite{Emery_2005}.}
  \begin{center}
    \begin{tabular}{l|c c c|c c c}
                                                                                                   & \multicolumn{3}{c|}{\textit{a}(\AA)} & \multicolumn{3}{c}{\textit{c}(\AA)}  \\
	  \hline
                                                                                                   & LDA      & PW91    & Expt.          & LDA      & PW91      & Expt.           \\
          CaC$_{6}$                                                                      &              &              &                   &              &               &                 \\
	  \hspace{1.5 mm} $\alpha$-stacking                             & 1.435    & 1.445   & N/A             & 4.34      & 4.50        & N/A             \\
	  \hspace{1.5 mm} $\alpha \beta$-stacking                & 1.435    & 1.450   & N/A             & 4.34      & 4.47        & N/A             \\
	  \hspace{1.5 mm} $\alpha \beta \gamma$-stacking & 1.434    & 1.450   & 1.444          & 4.35      & 4.50       & 4.524           \\
	  \hline
	  CaC$_{8}$ $\alpha$-stacking                                         & 1.435    & 1.445   & N/A              & 4.31      & 4.45       & N/A             \\
	  \hline
	  CaC$_{14}$ $\alpha$-stacking                                       & 1.424    & 1.437   & N/A              & 4.28      & 4.45       & N/A             \\
    \end{tabular}
    \label{tab:CaCncellparameters}
  \end{center}
\end{table*}

\subsection{H$_{2}$ in CaC$_{n}$}
\label{subsec:H2 in CaCn}

Hydrogen was introduced into the CaC$_{n}$ structures, with $n=6,8,14$, to test the effect of metal density on the binding energy of H. By choosing only the the three stoichiometries mentioned we have limited the search for the equilibrium H position to be only within \emph{two} inequivalent hexagons: one with the Ca atom in and one without. These are the only two distinct spaces that exist within CaC$_{6}$, CaC$_{8}$ and CaC$_{14}$ since every empty hexagon is equivalent. 

In order to sample these two spaces fully it was decided that hydrogen should be imported in both molecular form as H$_{2}$ and atomically as 2H. Six distinct starting positions, shown in figure \ref{fig:H+H+CaC6_start_posn}, were chosen for hydrogen with positions (\emph{a}) to (\emph{d}) being pairs of H atoms; and positions (\emph{e}) and (\emph{f}) being molecular H$_{2}$, with all centres of mass lying perfectly mid-way between the graphite sheets. It should be noted that position (\emph{f}) is perpendicular and position (\emph{e}) is parallel to the graphite sheets. This results in six different atomic relaxations. Before these could be performed the hydrogenated CaC$_{n}$ unit cell was relaxed with all atomic positions fixed in the starting position. Once the equilibrium starting structure was found an atomic relaxation was then started using the automatic VASP algorithm. A plane-wave basis set energy cutoff of 800~eV was used and a $k$-point grid with spacings always smaller than 0.04~\AA$^{-1}$ were used for all calculations. During relaxations (\emph{b}) and (\emph{c}) both relaxed back to position (\emph{a}) showing the presence of a strong local minimum. 

The hydrogen bonding energies were calculated by subtracting the free hydrogen molecule energy and the relaxed CaC$_{n}$ energy from the hydrogenated CaC$_{n}$, i.e. 
\begin{multline}
	E\left({\mathrm{H}}_{2}\;{\mathrm{binding\;energy}}\right) = \\ 
	E\left({\mathrm{H}}_{2}\;{\mathrm{in}}\;{\mathrm{CaC}}_{n}\right) - \left[E\left({\mathrm{H}}_{2}\right) + E\left({\mathrm{CaC}}_{n}\right)\right] \nonumber,
\end{multline}
thus giving a negative energy for favourable absorption of H$_{2}$.

The results in table \ref{tab:CaCnDifference} show the absorption energies for hydrogen in three different stoichiometries of CaC$_{n}$ in the most favourable position. In CaC$_{6}$ position (\emph{a}), two individual H atoms produced the highest binding energy. For CaC$_{14}$ position (\emph{f}), molecular H$_{2}$ and perpendicular to the graphite sheets produced the highest binding energy. In CaC$_{8}$ the most favourable absorption was of an H$_{2}$ molecule, with a negligible difference (approximately 3~meV) in binding energy between positions (\emph{e}) (parallel to the graphite sheets) and position (\emph{f}) (perpendicular to the graphite sheets). 
 
\begin{table}[htp]
  \caption{Calculated absorption energies for hydrogen in position (\emph{a}) (two individual H atoms) for CaC$_{6}$; and position (\emph{f}) (molecular H$_{2}$ perpendicular to the graphite sheets) for CaC$_{8}$ and CaC$_{14}$.}
  \begin{center}
    \begin{tabular}{l|c c}
                  & \multicolumn{2}{c}{H$_{2}$ Binding Energies [eV]} \\
                  & LDA     & PW91  \\
      \hline
      CaC$_{6}$   & -0.25   & -0.24 \\
      CaC$_{8}$   & -0.08   &  0.31 \\
      CaC$_{14}$  & -0.01   &  0.30 \\
    \end{tabular}
    \label{tab:CaCnDifference}
  \end{center}
\end{table}

It is interesting to compare this with hydrogen absorption in KC$_{8}$ GIC. For that material, the lowest energy configuration, by inelastic neutron scattering, is found to be one in which the H remains molecular and resides in a position near (\emph{f}) \cite{Lovell_2008}. \citet{Lovell_2008} perform DFT calculations to find that the difference in H$_{2}$ binding energy between the perpendicular (\emph{f}) and the parallel (\emph{e}) configuration in KC$_{8}$ is just a few meV, in qualitative agreement with the predictions given here for CaC$_{8}$.  

As shown in table \ref{tab:CaCnDifference} hydrogen was favourably absorbed with an LDA binding energy of $-0.25$~eV in CaC$_{6}$. It is worth noting that when the relaxations were performed using the same plane wave cut off and k-point grid as \citet{Cobian_2008} the results were in quantitative agreement with \citet{Cobian_2008}. The hydrogen was favourably absorbed with PW91 also, with a binding energy of $-0.24$~eV. The C-C bond length relaxed to $a=1.407$~\AA, a modest $-2.0\%$ decrease upon pure CaC$_{6}$. The inter-planar separation changed to $c=5.24$~\AA $\;$  upon loading of hydrogen; a $20\%$ increase, as expected from the formation of a Ca-H bond and thus the weakening of the Ca-C bond as noted by \citet{Cobian_2008}.

\subsection{Calcium hydride (CaH$_{2}$) crystal}
\label{subsec:CaH2}

The equilibrium crystal parameters presented in this work are compared with the experimental results from \citet{Wu_2007} in table \ref{CaH2parameters}. The calculated structure of calcium hydride also agrees well with the calculated structure of \citet{Li_2007} and \citet{Wolverton_2007}. The agreement between calculated and experimental crystal parameters is very good in PW91, and slightly worse
with LDA. 
 
\begin{table}[htp]
  \caption{Wyckoff 4\emph{c} coordinates within the CaH$_{2}$ unit cell. Note that the $x$-, $y$- and $z$-coordinates are fractional coordinates of the lattice vectors.}
  \begin{center}
    \begin{tabular}{c|c c c}
      Position & $x$-coord. & $y$-coord. & $z$-coord. \\
      \hline
      1 & $x$              & $0.25$ & $z$ \\
      2 & $-x+\frac{1}{2}$ & $0.75$ & $z+\frac{1}{2}$ \\
      3 & $-x$             & $0.75$ & $-z$ \\
      4 & $x+\frac{1}{2}$  & $0.25$ & $-z+\frac{1}{2}$ \\
    \end{tabular}
    \label{tab:WyckoffPosns}
  \end{center}
\end{table}

\begin{table*}[htp]
  \caption{Calculated unit cell parameters and atomic positions for the CaH$_{2}$ crystal and their comparison with experiment.}
  \begin{center}
    \begin{tabular}{l|c c| c c}
                                            & \multicolumn{2}{l|}{Calculated 0 K}                         & \multicolumn{2}{l}{Experimental Ref. \cite{Wu_2007}} \\ 
                                            & \multicolumn{2}{l|}{(without zero point energies)}  & & \\
                                            & LDA            & PW91               & 9 K          & 298 K             \\
      \hline    
      Unit Cell                             &                &                    &              &                   \\
      \hspace{3 mm} Parameters              &                &                    &              &                   \\
      \hspace{3 mm} \textit{a} (\AA)        & 5.72168        & 5.90763            & 5.92852(5)   & 5.94753(6)        \\
      \hspace{3 mm} \textit{b} (\AA)        & 3.46614        & 3.57014            & 3.57774(3)   & 3.59326(3)        \\
      \hspace{3 mm} \textit{c} (\AA)        & 6.56912        & 6.77002            & 6.78956(6)   & 6.80185(7)        \\
      \hspace{3 mm} \textit{V} (\AA$^{3}$)  & 130.280        & 142.787            & 144.011(1)   & 145.362(2)        \\
      \hline
      Ca1 (4\textit{c})                     &                &                    &              &                   \\
      \hspace{3 mm} $x$                     & 0.23898        & 0.23928            & 0.2387(1)    & 0.2397(2)         \\
      \hspace{3 mm} $z$                     & 0.11180        & 0.10988            & 0.1102(1)    & 0.1093(1)         \\
      \hline
      H1 (Calc.)/                           &                &                    &              &                   \\
      \hspace{3 mm} D1 (Exp.)(4\textit{c})  &                &                    &              &                   \\
      \hspace{3 mm} $x$                     & 0.35479        & 0.35586            & 0.3558(1)    & 0.3551(1)         \\ 
      \hspace{3 mm} $z$                     & 0.42843        & 0.42718            & 0.4276(1)    & 0.4268(1)         \\
      \hline
      H2 (Calc.)/                           &                &                    &              &                   \\
      \hspace{3 mm} D2 (Exp.)(4\textit{c})  &                &                    &              &                   \\
      \hspace{3 mm} $x$                     & 0.97416        & 0.97482            & 0.9750(1)    & 0.9743(2)         \\ 
      \hspace{3 mm} $z$                     & 0.68139        & 0.67648            & 0.6756(1)    & 0.6759(1)         \\
    \end{tabular}
    \label{CaH2parameters}
  \end{center}
\end{table*}

To ascertain the stability of the loaded H$_{2}$ $+$ CaC$_{n}$ system an energy comparison with CaH$_{2}$ and graphite was performed. The stability was established by subtracting the energy of the H$_{2}$ $+$ CaC$_{n}$ system from the separate energies of the calcium hydride crystal and graphite, i.e.
\begin{multline}
  \Delta{E} = \left[E\left({\mathrm{CaH}}_{2}\right) + E\left({\mathrm{C}}_{n}\right)\right] \\
	 - E\left({\mathrm{H}}_{2}\;{\mathrm{in}}\;{\mathrm{CaC}}_{n}\right),
\end{multline}
which gives $\Delta{E}$ to be negative if the H$_{2}$ $+$ CaC$_{n}$ system is unstable against decomposition into the metal hydride and pure graphite. The calculated values of $\Delta{E}$ are reported in table \ref{tab:stability}. We see that both LDA and PW91 predict that all the loaded H$_{2}$ $+$ CaC$_{n}$ structures are strongly unstable with respect to decomposition into graphite plus CaH$_{2}$.

\begin{table}[htp]
  \caption{Calculated stability energies, $\Delta{E}$, against decomposition.}
    \begin{center}
    \begin{tabular}{l|c c}
	& \multicolumn{2}{c}{$\Delta{E}$ [eV] for} \\
	& \multicolumn{2}{c}{H$_{2}$ $+$ CaC$_{n}$ $\rightarrow$ CaH$_{2}$ $+$ C$_{6}$}  \\
                                 & LDA       & PW91  \\
      \hline
      $\Delta{E}$ for CaC$_{6}$   & -1.11     & -1.15  \\
      $\Delta{E}$ for CaC$_{8}$   & -1.38     & -1.69  \\
      $\Delta{E}$ for CaC$_{14}$  & -1.81     & -1.74  \\
    \end{tabular}
    \label{tab:stability}
  \end{center}
\end{table}

\section{Discussion and conclusions}  %
\label{sec:conclusion}                %

Our calculations completely confirm earlier predictions that the intercalated graphitic material CaC$_{6}$, does favourably absorb hydrogen, with an absorption energy within the U.S. Department of Energy range of $0.2-0.8$~eV. However, we have also found that the hydrogen-loaded CaC$_{n}$ material is strongly unstable with respect to decomposition into CaH$_{2}$ crystal plus graphite, the energy release associated with this decomposition being $1.1-1.8$~eV. This explains recent experiments \cite{Srinivas_2009}, which showed that attempted hydrogenation of CaC$_{6}$ prompted the complete deintercalation of the starting material to leave just CaH$_{2}$ and graphite.

In considering the reliability of our predictions, we must, of course, consider the likely size of errors with our DFT calculations. It is well known that DFT can easily suffer from errors of a few tenths of an eV in cohesive energies. We have seen in table \ref{tab:stability} that the values of energy release for the decomposition of hydrogenated CaC$_{8}$ from LDA and PW91 differ by $\sim$0.3~eV. A recent study of the energetics of MgH$_{2}$ by \citet{Pozzo_2008} showed that DFT errors in the cohesive energy can be as large as 0.5~eV, but that LDA and PW91 predictions bracket the true value. However even with errors of this size, our calculations leave no doubt that the decomposition is strongly exothermic, with a decomposition energy of at least 1.0~eV.

It is worth noting that thermodynamic instability with respect to decomposition does not necessarily tell us that decomposition will occur spontaneously at room temperature, because the decomposition might conceivably be kinetically hindered. However, the experiments \cite{Srinivas_2009} show that at room temperature and within one hour of hydrogen exposure the sample lost all evidence of the CaC$_{6}$ structure, and within five hours showed only the CaH$_{2}$ and graphite structures. These time-scales suggest that the rate-limiting step probably has an activation energy of $\sim$0.9~eV. The computational search for possible decomposition pathways would be a major undertaking, and is well beyond the scope of the present work. 

The results we have found raise the interesting question of what determines whether a hydrogenated metal GIC will be unstable or not with respect to decomposition. We know that hydrogenated Li-GIC and Ca-GIC are both unstable. However, K-GIC appears to be stable, and a number of experiments have been reported on the hydrogenated material \cite{Lovell_2006, Lovell_2008, Cheng_2001a, Lagrange_1987}. We suggest that the dominant factor may be the cohesive energies of the metal hydride, with those having large cohesive energies being unstable. This would probably fit the facts, because the larger lattice parameter of KH compared with LiH should give it a smaller cohesive energy. The double charge of Ca in CaH$_{2}$ should also make its cohesive energy larger than that of KH. This is an interesting question for future work. 

\section{Acknowledgements}   %
We would like to thank Jorge {\'{I}}{\~{n}}iguez and Eduardo Hernandez for their extremely helpful correspondence during this work. Charlie Wood wishes to thank Rich Spinney and Arthur Lovell for useful discussions. This work was funded by the EPSRC. 

\bibliographystyle{model1a-num-names}
\bibliography{bib}

\begin{thebibliography}{22}
\expandafter\ifx\csname natexlab\endcsname\relax\def\natexlab#1{#1}\fi
\providecommand{\bibinfo}[2]{#2}
\ifx\xfnm\relax \def\xfnm[#1]{\unskip,\space#1}\fi
\bibitem[{Mehta and Cooper(2003)}]{Mehta_2003}
\bibinfo{author}{V.~Mehta}, \bibinfo{author}{J.~S. Cooper},
  \bibinfo{journal}{J. Power Sources} \bibinfo{volume}{114}
  (\bibinfo{year}{2003}) \bibinfo{pages}{32--53}.
\bibitem[{Zhou(2005)}]{Zhou_2005}
\bibinfo{author}{L.~Zhou}, \bibinfo{journal}{Renew. Sust. Energ. Rev.}
  \bibinfo{volume}{9} (\bibinfo{year}{2005}) \bibinfo{pages}{395--408}.
\bibitem[{Harris et~al.(2004)Harris, Book, Anderson, and Edwards}]{Harris_2004}
\bibinfo{author}{R.~Harris}, \bibinfo{author}{D.~Book},
  \bibinfo{author}{P.~Anderson}, \bibinfo{author}{P.~Edwards},
  \bibinfo{journal}{The Fuel Cell Review} \bibinfo{volume}{2004}
  (\bibinfo{year}{2004}) \bibinfo{pages}{17--22}.
\bibitem[{Ross(2006)}]{Ross_2006}
\bibinfo{author}{D.~K. Ross}, \bibinfo{journal}{Vacuum} \bibinfo{volume}{80}
  (\bibinfo{year}{2006}) \bibinfo{pages}{1084--1089}.
\bibitem[{DoE(2009)}]{DoE_2009}
\bibinfo{title}{Hydrogen, {F}uel {C}ells {\&} {I}nfrastructure {T}echnologies
  {P}rogram, {M}ulti-{Y}ear {R}esearch, {D}evelopment and {D}emonstration
  {P}lan}, \bibinfo{year}{2009}.
\bibitem[{Dresselhaus and Dresselhaus(1980)}]{Dresselhaus_1980}
\bibinfo{author}{S.~Dresselhaus}, \bibinfo{author}{G.~Dresselhaus},
  \bibinfo{journal}{Adv. Phys.} \bibinfo{volume}{51} (\bibinfo{year}{1980})
  \bibinfo{pages}{1--186}.
\bibitem[{Lovell et~al.(2006)Lovell, Bennington, Skipper, Gejke, Thompson, and
  Adams}]{Lovell_2006}
\bibinfo{author}{A.~Lovell}, \bibinfo{author}{S.~M. Bennington},
  \bibinfo{author}{N.~T. Skipper}, \bibinfo{author}{C.~Gejke},
  \bibinfo{author}{H.~Thompson}, \bibinfo{author}{M.~A. Adams},
  \bibinfo{journal}{Physica B} \bibinfo{volume}{385-386} (\bibinfo{year}{2006})
  \bibinfo{pages}{163--165}.
\bibitem[{Lovell et~al.(2008)Lovell, Fernandez-Alonso, Skipper, and
  Refson}]{Lovell_2008}
\bibinfo{author}{A.~Lovell}, \bibinfo{author}{F.~Fernandez-Alonso},
  \bibinfo{author}{N.~T. Skipper}, \bibinfo{author}{F.~Refson},
  \bibinfo{journal}{Phys. Rev. Lett.} \bibinfo{volume}{101}
  (\bibinfo{year}{2008}) \bibinfo{pages}{126101}.
\bibitem[{Carlile et~al.(1998)Carlile, Kearley, Lindsell, and
  White}]{Carlile_1998}
\bibinfo{author}{C.~J. Carlile}, \bibinfo{author}{G.~J. Kearley},
  \bibinfo{author}{G.~Lindsell}, \bibinfo{author}{J.~White},
  \bibinfo{journal}{Physica B} \bibinfo{volume}{341-243} (\bibinfo{year}{1998})
  \bibinfo{pages}{491--494}.
\bibitem[{Beaufils et~al.(1981)Beaufils, Crowley, Rayment, Thomas, and
  White}]{Beaufils_1981}
\bibinfo{author}{J.~P. Beaufils}, \bibinfo{author}{T.~Crowley},
  \bibinfo{author}{T.~Rayment}, \bibinfo{author}{R.~Thomas},
  \bibinfo{author}{J.~W. White}, \bibinfo{journal}{Mol. Phys.}
  \bibinfo{volume}{44} (\bibinfo{year}{1981}) \bibinfo{pages}{1297--1269}.
\bibitem[{Cobian and \'{I}\~{n}iguez(2008)}]{Cobian_2008}
\bibinfo{author}{M.~Cobian}, \bibinfo{author}{J.~\'{I}\~{n}iguez},
  \bibinfo{journal}{J. Phys.-Condens. Matter} \bibinfo{volume}{2}
  (\bibinfo{year}{2008}) \bibinfo{pages}{285212}.
\bibitem[{Srinivas et~al.(2009)Srinivas, Howard, Bennington, Skipper, and
  Ellerby}]{Srinivas_2009}
\bibinfo{author}{G.~Srinivas}, \bibinfo{author}{C.~A. Howard},
  \bibinfo{author}{S.~M. Bennington}, \bibinfo{author}{N.~T. Skipper},
  \bibinfo{author}{M.~Ellerby}, \bibinfo{journal}{J. Mater. Chem.}
  \bibinfo{volume}{19} (\bibinfo{year}{2009}) \bibinfo{pages}{5239--5243}.
\bibitem[{Enoki et~al.(1990)Enoki, Miyajima, Sano, and Inokuchi}]{Enoki_1990}
\bibinfo{author}{T.~Enoki}, \bibinfo{author}{S.~Miyajima},
  \bibinfo{author}{M.~Sano}, \bibinfo{author}{H.~Inokuchi},
  \bibinfo{journal}{Journal of Materials Research} \bibinfo{volume}{5}
  (\bibinfo{year}{1990}) \bibinfo{pages}{435--466}.
\bibitem[{Shi et~al.(1997)Shi, Barker, Sa{\"{\i}}di, Koksbang, and
  Morris}]{Shi_1997}
\bibinfo{author}{H.~Shi}, \bibinfo{author}{J.~Barker}, \bibinfo{author}{M.~Y.
  Sa{\"{\i}}di}, \bibinfo{author}{R.~Koksbang}, \bibinfo{author}{L.~Morris},
  \bibinfo{journal}{J. Power Sources} \bibinfo{volume}{68}
  (\bibinfo{year}{1997}) \bibinfo{pages}{291--295}.
\bibitem[{Emery et~al.(2005)Emery, H{\'{e}}rald, and Lagrange}]{Emery_2005}
\bibinfo{author}{N.~Emery}, \bibinfo{author}{C.~H{\'{e}}rald},
  \bibinfo{author}{P.~Lagrange}, \bibinfo{journal}{J. Solid State Chem.}
  \bibinfo{volume}{178} (\bibinfo{year}{2005}) \bibinfo{pages}{2947--2952}.
\bibitem[{Srinivas et~al.(2009)Srinivas, Howard, Skipper, Bennington, and
  Ellerby}]{Srinivas_2009a}
\bibinfo{author}{G.~Srinivas}, \bibinfo{author}{C.~Howard},
  \bibinfo{author}{N.~Skipper}, \bibinfo{author}{S.~Bennington},
  \bibinfo{author}{M.~Ellerby}, \bibinfo{journal}{Physica C: Superconductivity}
  \bibinfo{volume}{469} (\bibinfo{year}{2009}) \bibinfo{pages}{2000 -- 2002}.
\bibitem[{Wu et~al.(2007)Wu, Zhou, Udovic, Rush, and Yildirim}]{Wu_2007}
\bibinfo{author}{H.~Wu}, \bibinfo{author}{W.~Zhou},
  \bibinfo{author}{T.~Udovic}, \bibinfo{author}{J.~Rush},
  \bibinfo{author}{T.~Yildirim}, \bibinfo{journal}{J. Alloy. Compd.}
  \bibinfo{volume}{436} (\bibinfo{year}{2007}) \bibinfo{pages}{51--55}.
\bibitem[{Li et~al.(2007)Li, Li, Yang, Cui, Ma, and Zou}]{Li_2007}
\bibinfo{author}{B.~Li}, \bibinfo{author}{Y.~Li}, \bibinfo{author}{K.~Yang},
  \bibinfo{author}{Q.~Cui}, \bibinfo{author}{Y.~Ma}, \bibinfo{author}{G.~Zou},
  \bibinfo{journal}{J. Phys.-Condens. Matter} \bibinfo{volume}{19}
  (\bibinfo{year}{2007}) \bibinfo{pages}{226205}.
\bibitem[{Wolverton and Ozolins(2007)}]{Wolverton_2007}
\bibinfo{author}{C.~Wolverton}, \bibinfo{author}{V.~Ozolins},
  \bibinfo{journal}{Phys. Rev. B} \bibinfo{volume}{75} (\bibinfo{year}{2007})
  \bibinfo{pages}{064101}.
\bibitem[{Pozzo and Alf{\`e}(2008)}]{Pozzo_2008}
\bibinfo{author}{M.~Pozzo}, \bibinfo{author}{D.~Alf{\`e}},
  \bibinfo{journal}{Phys. Rev. B} \bibinfo{volume}{77} (\bibinfo{year}{2008})
  \bibinfo{pages}{104103}.
\bibitem[{Cheng et~al.(2001)Cheng, Pez, Kern, Kresse, and Hafner}]{Cheng_2001a}
\bibinfo{author}{H.~Cheng}, \bibinfo{author}{G.~Pez},
  \bibinfo{author}{G.~Kern}, \bibinfo{author}{G.~Kresse},
  \bibinfo{author}{J.~Hafner}, \bibinfo{journal}{The Journal of Physical
  Chemistry B} \bibinfo{volume}{105} (\bibinfo{year}{2001})
  \bibinfo{pages}{736--742}.
\bibitem[{Lagrange et~al.(1987)Lagrange, Guerard, Mareche, and
  Herold}]{Lagrange_1987}
\bibinfo{author}{P.~Lagrange}, \bibinfo{author}{D.~Guerard},
  \bibinfo{author}{J.~F. Mareche}, \bibinfo{author}{A.~Herold},
  \bibinfo{journal}{Journal of the Less Common Metals} \bibinfo{volume}{131}
  (\bibinfo{year}{1987}) \bibinfo{pages}{371 -- 378}.

\end{thebibliography}

\end{document}